\documentclass[journal]{IEEEtran}
\usepackage[noadjust]{cite}
\pdfoutput = 1

\ifCLASSINFOpdf
\usepackage[pdftex]{graphicx}
\else
\usepackage[dvips]{graphicx}
\fi

\usepackage{amsmath}
\usepackage{array}

\ifCLASSOPTIONcompsoc
  \usepackage[caption=false,font=normalsize,labelfont=sf,textfont=sf]{subfig}
\else
  \usepackage[caption=false,font=footnotesize]{subfig}
\fi
\usepackage{url}
\usepackage{amssymb}

\DeclareMathOperator*{\argmin}{arg\, min}

\usepackage{algorithm}
\usepackage{algorithmicx}
\usepackage[noend]{algpseudocode}
\newcommand*\Let[2]{\State #1 $\gets$ #2}
\algrenewcommand\algorithmicrequire{\textbf{Input:}}
\algrenewcommand\algorithmicensure{\textbf{Output:}}

\usepackage[normalem]{ulem}
\usepackage{tikz}
\usepackage{color}
\definecolor{mygreen}{rgb}{0 0.7 0}
\definecolor{mygray}{rgb}{0.6 0.6 0.6}
\definecolor{myblue}{rgb}{0.7 0.7 1}
\definecolor{mypink}{rgb}{1 0 1}

\DeclareMathOperator*{\argmax}{arg\,max}

\usepackage{multirow}
\usepackage{mathtools}
\usepackage{makecell}
\usepackage{rotating}
\def\block(#1,#2)#3{\multicolumn{#2}{c}{\multirow{#1}{*}{$ #3 $}}}

\begin{document}

\title{\huge{Multi-fidelity sensor selection:  Greedy algorithms to place cheap and expensive sensors with cost constraints }}

\author{Emily~Clark\thanks{E. Clark is with the Department of Physics at
the University of Washington, Seattle, WA, 98195-1560 USA email:
eclark7@uw.edu},
        Steven~L.~Brunton\thanks{S. Brunton is with the Department of Mechanical Engineering
at the University of Washington, Seattle, WA, 98195-2600 USA email:
sbrunton@uw.edu},~\IEEEmembership{Senior Member,~IEEE,}
        J.~Nathan~Kutz\thanks{J. N. Kutz is with the Department of Applied
Mathematics at the University of Washington, Seattle, WA, 98195-3925 
USA email: kutz@uw.edu},~\IEEEmembership{Member,~IEEE}}

\markboth{arXiv Submission}%
{Clark \MakeLowercase{\textit{et al.}}: Sensor Placement with Cost Constraints}

\maketitle

\begin{abstract}
We develop greedy algorithms to approximate the optimal solution to the multi-fidelity sensor selection problem, which is a cost constrained optimization problem prescribing the placement and number of cheap (low signal-to-noise) and expensive (high signal-to-noise) sensors in an environment or state space.  
Specifically, we evaluate the composition of cheap and expensive sensors, along with their placement, required to achieve accurate reconstruction of a high-dimensional state.
We use the column-pivoted QR decomposition to obtain preliminary sensor positions.   
How many of each type of sensor to use is highly dependent upon the sensor noise levels, sensor costs, overall cost budget, and the singular value spectrum of the data measured.  
Such nuances allow us to provide sensor selection recommendations based on computational results for asymptotic regions of parameter space. 
We also present a systematic exploration of the effects of the number of modes and sensors on reconstruction error when using one type of sensor.  
Our extensive exploration of multi-fidelity sensor composition as a function of data characteristics is the first of its kind to provide guidelines towards optimal multi-fidelity sensor selection.
\end{abstract}


%



\section{Introduction}
\label{sec:intro}

\IEEEPARstart{S}{parse} sensor selection, which is the underlying mathematical theory that optimizes the placement of a limited number of sensors in an environment or state space, is a combinatorially hard problem.  
Obtaining tractable approximate solutions is essential to many fields in the engineering and physical sciences, including scientific experiments~\cite{antchev2010totem, taylor2004towards}, reduced-order modeling~\cite{willcox2006unsteady, chaturantabut2010nonlinear}, robotics~\cite{koshizen2000improved,mahoney2016inseparable}, industry~\cite{samadiani2012reduced, liu2015entropy, chua2015sensor,Manohar2018jms}, public works~\cite{o2007smartcoast, raut2013intelligent}, and climate science~\cite{zhang2016simultaneous, dolan2007harmful}. One facet of the sensor selection problem rarely addressed is the optimal placement of multiple types of sensors that can have significantly different costs and performance metrics. Sensors often measure different quantities (multi-modal sensing), or they could measure the same quantity but with different qualities (multi-fidelity sensing), i.e. different noise level, bias, sampling rate, range, power consumption, construction cost, placement cost, or longevity. Given access to two or more types of sensor, what principled mathematical method can determine how many of each type to select? Where should they be placed in order to optimize relevant factors such as performance and cost? This work considers the sparse selection of multi-fidelity sensors with each having significantly different costs and signal-to-noise levels. Specifically, our proposed sparse selection procedure is aimed at determining the optimal number of each sensor type and their placement so as to simultaneously minimize reconstruction error and financial cost. We also address the optimal number of sensors relative to the rank of the data set, and the basis of choice for a given sensor type.

Because of the wide range of sparse sensing applications, there is a significant body of theoretical work on principled sensor selection spanning many disciplines.  
In many cases, the principled methods are framed as a mathematical technique for interpolating a high-dimensional state space from a small number of measurements.  Methods with this interpolation viewpoint include gappy proper orthogonal decomposition~\cite{willcox2006unsteady}, placing sensors at the maxima of singular vectors~\cite{yildirim2009efficient, yang2010eof}, leveraging multiscale physics~\cite{manohar2019optimized},  interpolating via asynchronous time-sampling~\cite{bright2016classification}, and nonlinear model order reduction methods, including EIM~\cite{barrault2004empirical}, DEIM~\cite{chaturantabut2010nonlinear}, and Q-DEIM~\cite{drmac2016new, manohar2018data}. Some sensor performance metrics are submodular and admit greedy sensor selection methods with performance guarantees, e.g.~\cite{krause2008near, shamaiah2010greedy, ranieri2014near}.
However, there is typically a disconnect between mathematical methods and practical implementations of sensor selection. Specifically, it is usual to assume one type of sensor, idealized conditions, and the availability of full-state training data, while neglecting important factors like sensor type, noise, and cost. Bridging the gap between idealized models and real-world practicalities is currently an open problem.

One important practical consideration is sensor cost, though certain submodular methods, such as~\cite{krause2006near, krause2007near}, can be modified to account for a nonuniform cost on sensor location, and Clark et al.~\cite{clark2018greedy} modify the column-pivoted QR decomposition to do the same. 
Sensor cost is also closely tied to the total number of sensors. 
The most common convention is to take an SVD basis and to use the same number of sensors as low-rank modes that approximately span the subspace where the data is embedded.
However, Peherstorfer et al~\cite{peherstorfer2018stabilizing} show that this leads to noise amplification, causing reconstruction errors to increase as more sensors are added. This can be mitigated by oversampling, taking more sensors than SVD modes. In~\cite{clark2018greedy}, the authors avoid noise amplification by instead using a randomized basis and, following the guidelines of~\cite{liberty2007randomized, halko2011finding}, perform undersampling, i.e. take more modes than sensors. But there is no systematic study of the effect of basis choice and over- and undersampling on reconstruction quality, so part of this work's contribution is to undertake such a study. We determine which basis has the lowest reconstruction error with a given number of sensors, and at what number of modes this minimum error occurs. We find that the SVD and randomized bases have similar minimum errors, but the randomized basis requires as many modes as possible for best performance, while the SVD basis achieves its lowest error with a moderate amount of oversampling.

Importantly, multi-fidelity sensor placement has received little consideration, and this is a central aim of our study.  The goal for multi-fidelity sensor placement is to judiciously use expensive, high-performance sensors in combination with cheap, low-performance sensors to provide accurate state-space reconstructions as cheaply as possible.  For us, the difference between cheap and expensive sensors will be associated with the additive noise at measurement.  Cheap sensors have large additive noise, and expensive sensors have low additive noise.  Note that Lahat et al~\cite{lahat2015multimodal} advocate for the benefits of multimodality for state estimation, but aside from the structure monitoring community, e.g.~\cite{zhang2011optimal, zhu2013multi, lin2018structural, zhang2016optimal, soman2014multi}, to our knowledge there is no principled method in the literature for the optimal selection of multiple types of sensors. In particular, there is little information on the placement of sensors with differing noise levels, though Kammer~\cite{kammer1992effects} describes a method for sensor selection in the case of location-dependent noise levels.

The optimization procedure can be stated colloquially as follows:  Is it better to purchase a large number of cheap sensors, a small number of expensive sensors, or a mix of both? In the latter case, how many of each type should be used, and where should they be placed? 
Alternatively, someone may already have sensors with low signal-to-noise levels, and needs to determine whether it is worth the time and money to design sensors with higher signal-to-noise levels. Quantitative answers to these questions can not only address feasibility issues, but can potentially save manufacturers significant amounts of money.   
However, the multi-fidelity  sparse sensor selection problem is complicated and nuanced, depending on the rank of the data considered, the relative and absolute costs and noise levels of the sensors, and the available budget.

In this work, we characterize optimal sensor selection with two types of sensors in a few asymptotic regimes of very low or very high costs and noise levels, for systems whose data is of low, medium, and high rank, given a fixed budget. We present empirical results and use them to formulate an initial set of guidelines for selecting a sensor type.  Broadly, if both types of sensors have low noise levels, it is usually better to use a larger number of cheap sensors. When the cheap sensors have a much higher noise level than the expensive sensors, a small number of expensive sensors tends to perform better. And when both sensors have similar, high noise levels, the rank of the system is the determining factor, with high-rank systems favoring a larger number of cheap sensors, and low-rank systems preferring a smaller number of expensive sensors.  Thus to evaluate the multi-fidelity sparse sensor selection problem, the nature of the measurement data plays a significant role.
To our knowledge, this is the first systematic study of the role of multi-fidelity, cost constrained sensor selection as a function of the data, cost, and noise.

The rest of the paper is organized as follows: Section~\ref{sec:methods} sets up the sensor selection problem and describes the column-pivoted QR algorithm to solve it. In Section~\ref{sec:NumSensorsModes} we explore the effects of basis choice and over- and undersampling with one type of sensor. Section~\ref{sec:Multi-fidelity} describes the multi-fidelity sensor selection problem and presents results and general guidelines from several asymptotic regimes. Finally, the discussion and conclusions are in Section~\ref{sec:Conclusion}.

\section{Sparse sensor selection}
\label{sec:methods}
We begin by describing the sparse sensing problem and the QR-based solution we adopt for our analysis. We make only small modifications in the case of multi-fidelity noisy sensors.

\subsection{Problem formulation}
\label{sec:setup}
We consider a high-dimensional system that would be difficult or impossible to measure fully. We wish to find the optimal sampling points that yield the best possible reconstructions of the full state.
We assume we have access to high-fidelity training data, either from a high-precision simulation or an experiment with full sensing.  Gathering the training data, which is a one-time expense, will be used to learn the optimal sensing locations. Once we have collected $m$ full-state snapshots ${\bf x}_i\in\mathbb{R}^n$, we gather them into a data matrix
\begin{align}
{\bf X}^{tr} = \left[\begin{array}{cccc}
{\bf x}_1 & {\bf x}_2 & \cdots & {\bf x}_m
\end{array}\right].
\end{align}
With this training data, we construct a modal basis to represent the system, such that any snapshot ${\bf x}_i$ can be constructed from the decomposition
\begin{align}
{\bf x}_i &= {\bf\Psi}{\bf a}_i,
\end{align}
where ${\bf\Psi}\in\mathbb{R}^{n\times r}$ and ${\bf a}_i\in\mathbb{R}^r$. There is flexibility to choose the basis that is right for the system and application at hand, see e.g.~\cite{clark2020sensor}, but often ${\bf\Psi}$ is taken to be the first $r$ left singular vectors of ${\bf X}^{tr}$.  We assume $r<\min(m,n)$.

Let ${\bf X}^{te} \in\mathbb{R}^{n\times\ell}$ be a test set of $\ell$ new snapshots that we wish to downsample. Given $p\ll n$ sensors, we must select $p$ locations to measure, indexed by the set $\{ J\}$, where $|J| = p$ and for all indices $j\in\{J\}$, $1\le j\le n$.  We downsample using the selection matrix ${\bf C}_{\{J\}}\in\mathbb{R}^{p\times n}$, where the rows of ${\bf C}_{\{J\}}$ are the unit vectors ${\bf e}^T_{\{J\}}$. Thus the measurements ${\bf Y}_{\{J\}}\in\mathbb{R}^{p\times\ell}$ are given by
\begin{subequations}
\begin{align}
{\bf Y}_{\{J\}} &= {\bf C}_{\{J\}}{\bf X}^{te}\\
&= {\bf C}_{\{J\}}{\bf\Psi}{\bf a}^{te}\\
&= {\bf\Theta}_{\{J\}}{\bf a}^{te},
\end{align}
\end{subequations}
where ${\bf a}^{te}\in\mathbb{R}^{r\times\ell}$ are the coefficients of the test set in the ${\bf\Psi}$ basis, and ${\bf\Theta}_{\{J\}} = {\bf C}_{\{J\}}{\bf\Psi}$ is the measurement matrix, consisting of the rows of ${\bf\Psi}$ that are in $\{J\}$.

To perform full-state reconstruction given sparse measurements, we take the minimum norm least squares solution,
\begin{align}
\label{eq:recona}
{\bf \hat{a}}^{te}_{\{J\}} &= \left({\bf\Theta}_{\{J\}}\right)^\dagger{\bf Y}_{\{J\}}\\
\label{eq:reconx}
{\bf\hat{X}}^{te}_{\{J\}} &= {\bf\Psi}{\bf\hat{a}}^{te}_{\{J\}},
\end{align}
where $\left({\bf\Theta}_{\{J\}}\right)^\dagger$ is the pseudoinverse of ${\bf\Theta}_{\{J\}}$.
The fractional reconstruction error for the system given the measurements $\{J\}$ is
\begin{equation}
E_{\{J\}} = \frac{||{\bf X}^{te}-{\bf\hat{X}}^{te}_{\{J\}}||_F}{||{\bf X}^{te}||_F},
\label{eq:error}
\end{equation}
where $||\cdot ||_F$ is the Frobenius norm.
With this performance metric, the sparse sensor selection problem becomes
\begin{equation}
\argmin_{\{J\}} E_{\{J\}}.
\label{eq:optE}
\end{equation}
This is a combinatorially hard problem, and we employ a greedy solution, as described in the next subsection.

\subsection{QR decomposition}
\label{sec:QR}
Our greedy sensor selection method is the column-pivoted QR decomposition, as described in~\cite{drmac2016new}. The QR decomposition was introduced for the solution of least squares problems in 1965 by Businger and Golub~\cite{businger1965linear}, and in 2016, Drma{\v{c}} and Gugercin~\cite{drmac2016new} showed that the greedy determinant maximization of the column-pivoted QR algorithm is highly effective for sensor placement.
For a given matrix ${\bf V}\in\mathbb{R}^{r\times n}$, $r\le n$, the QR decomposition with column pivoting is given by:
\begin{equation}
\label{eq:QR}
{\bf V}{\bf P} = {\bf Q}{\bf R},
\end{equation}
where ${\bf Q}\in\mathbb{R}^{r\times r}$ is unitary, ${\bf R}\in\mathbb{R}^{r\times n}$ is upper triangular, and ${\bf P}\in\mathbb{R}^{n\times n}$ is a column permutation matrix.

An outline of the column-pivoted QR decomposition algorithm with Householder reflections is as follows, also see Algorithm~\ref{alg:qr} for full details: We initialize the permutation matrix to $\mathbb{I}_n$, select as the first pivot the column ${\bf v}_j$ of ${\bf V}$ with the largest norm, and swap it with the first column of ${\bf V}$ through the permutation matrix. We then apply a Householder transformation to ${\bf V}$ such that ${\bf v}_j$ becomes $\left(\begin{array}{cccc} ||{\bf v}_j|| & 0 & \cdots & 0\end{array}\right)^T$. We repeat the pivoting and the Householder reflection on the $(r-1)\times(n-1)$ lower right-hand submatrix of the transformed ${\bf V}$, yielding a $2\times2$ upper triangular submatrix in the upper left corner. Repeat the pivoting and reflecting for a total of $r$ times. In this way, the matrix ${\bf V}$ is iteratively transformed to be upper triangular, and because Householder reflections are unitary, the decomposition is achieved.

This algorithm imposes a diagonal dominance structure in ${\bf R}$, i.e.
\begin{equation}
|r_{ii}|^2 \ge \sum_{j=1}^k |r_{jk}|^2, \hspace{24pt} 1\le i\le k\le r,
\end{equation}
and therefore maximizes the determinant of ${\bf R}$. In fact, because the Householder transformations are unitary, this procedure maximizes the determinant of the upper left submatrix that has been made upper triangular at every iteration.
To perform sensor selection, we apply the column-pivoted QR algorithm to ${\bf\Psi}^*$ and take ${\bf C}_{\{J\}}^T = {\bf P}_{:,1:p}$, where ${\bf P}_{:,1:p}$ are the first $p$ columns of ${\bf P}$. In other words, the sensor locations correspond to the rows of ${\bf\Psi}$ with the iteratively largest norms, such that we maximize the determinant of those selected rows, ${\bf\Theta}_{\{J\}}$.

Thus, QR with column pivoting does not solve Eq.~\ref{eq:optE}, but rather it greedily solves
\begin{equation}
\argmax_{\{J\}} \left| \det {\bf\Theta}_{\{J\}} \right|.
\end{equation}
Maximizing the determinant maximizes the volume of the measurement matrix, ensuring that we are selecting informative locations. Moreover, this procedure causes ${\bf\Theta}_{\{J\}}$ to be stable under inversion by minimizing its condition number
\begin{equation}
K = \frac{\sigma_{max}}{\sigma_{min}},
\end{equation}
where $\sigma$ are the singular values of ${\bf\Theta}_{\{J\}}$. A large condition number leads to amplification of any noise in the signal when the full-state reconstruction is performed, so although we have not directly solved the optimization problem in terms of reconstruction error, we can be assured that the column-pivoted QR decomposition selects desirable sensor locations.
We highlight the work of 
Drma{\v{c}} and Gugercin~\cite{drmac2016new} who showed the algorithm to be highly effective for sensor placement.

\begin{algorithm}[t]
  \caption[Caption]{\label{alg:qr} QR pivoting.}
  \begin{algorithmic}[1]
    \Require{matrix ${\bf V}\in\mathbb{R}^{r\times n}$, number of sensors $p\le r$}
    \Ensure{Partial permutation matrix ${\bf P}$}
    \Let{${\bf P}$}{$\mathbb{I}_n$}
    \For{$k = 1,\ldots,p$}
    \Let{$j_k$}{$\arg\max\limits_{j\ge k}||{\bf V}_{k:r,j}||$}
    \State \texttt{Swap}$({\bf P}_{:,k},{\bf P}_{:,j_k})$.
    \State Calculate Householder reflection ${\bf Q}_k$ such that ${\bf Q}_k \left(\begin{array}{cccc}
	V_{k,j_k} & V_{k+1,j_k} & \cdots & V_{r,j_k}\end{array}\right)^T = \left(\begin{array}{cccc}
	||{\bf V}_{k:r,j_k}|| & 0 & \cdots & 0 \end{array}\right)^T$.
    \Let{${\bf V}$}{$\text{diag}(\mathbb{I}_{k-1},{\bf Q}_k){\bf V}{\bf P}$}
    \EndFor
    \Return{${\bf P}_{:,1:p}$}
  \end{algorithmic}
\end{algorithm}

\section{Effect of number of sensors and modes}
\label{sec:NumSensorsModes}
Before incorporating multi-fidelity measurements, we present a thorough exploration of the effect of the number of sensors and modes on reconstruction error, for two common basis choices.
The first is the basis of left singular vectors of ${\bf X}^{tr}$:
\begin{align}
{\bf X}^{tr} &= {\bf U}{\bf \Sigma}{\bf V}^*\\
\label{eq:SVDBasis}
{\bf\Psi} &= {\bf U}_{:,1:r},
\end{align}
where ${\bf U}$ and ${\bf V}$ are orthogonal, ${\bf\Sigma}$ is diagonal with entries $\sigma_1\ge\sigma_2\ge\dots\ge\sigma_n\ge0$, and ${\bf U}_{:,1:r}$ contains the first $r$ columns of ${\bf U}$. The is the default choice for most sparse sensing applications. The SVD provides the optimal low-rank approximation of a data set~\cite{eckart1936approximation}, and the SVD basis has proven highly effective for sparse sensing, see, for example,~\cite{manohar2018data, drmac2016new}. However, Peherstorfer, Drma{\v{c}}, and Gugercin~\cite{peherstorfer2018stabilizing} discover that the convention of taking the same number of modes and sensors in an SVD basis leads to noise amplification, with the result that reconstruction error actually increases with $p$. The authors prove that oversampling, taking more sensors (interpolation points) than modes, rectifies the problem, but while they explore oversampling methods, they do not go in-depth about the optimal degree of oversampling.

The second basis consists of randomized linear combinations of the columns of ${\bf X}^{tr}$, as described in, e.g.,~\cite{halko2011finding,liberty2007randomized,erichson2016randomized}:
\begin{equation}
\label{eq:RandBasis}
{\bf\Psi} = {\bf X}^{tr}{\bf G},
\end{equation}
where ${\bf G} \in\mathbb{R}^{m\times r}$ has i.i.d. Gaussian random entries. This basis is faster to calculate than the SVD, making it particularly useful for large data sets, and it is able to capture much of the system's energy with high statistical likelihood. The likelihood of randomized rank reduction sufficiently capturing the system's energy increases with the number of modes, where the number of modes should be greater than the rank of the system. In Clark et al~\cite{clark2018greedy}, the authors use a randomized reduced basis for sensor placement with undersampling, taking $r = 2p$. Doing so eliminates noise amplification and leads to good reconstructions, but again, the authors do not explore the optimal number of modes and sensors.

Here, we consider both the SVD and randomized bases, and vary the number of modes and sensors to determine the optimal amount of over- and undersampling for each basis. We also look at which basis has the best performance overall. We test on the Extended Yale Face Database B~\cite{cai2007learning, cai2007spectral, cai2006orthogonal, he2005face}, which consists of around 64 images each of 38 individuals under different lighting conditions, with the photographs resized to $32\times32$ pixels, for $n=1024$. There are a total of 2414 photographs, and the data has a slow singular value decay and a Gavish-Donoho cutoff~\cite{gavish2014optimal} around rank 350. We randomly select $80\%$ of the snapshots for ${\bf X}^{tr}$ and the rest of the photographs form the test set ${\bf X}^{te}$.

We select the number of modes $r$, the number of sensors $p$, and our basis, either Eq.~\ref{eq:SVDBasis} or~\ref{eq:RandBasis}, then use Algorithm~\ref{alg:qr} to place sensors. We let the sensors have $2\%$ signal-to-noise levels, such that the measurements are given by
\begin{equation}
{\bf Y} = {\bf C}_{\{J\}} {\bf X}^{te} + \boldsymbol{\epsilon},
\end{equation}
where $\boldsymbol{\epsilon}\in\mathbb{R}^{p\times \ell}$ has Gaussian i.i.d. entries with variance $2\%$ of the overall variance of the data set. We perform reconstruction as in Eq.~\ref{eq:recona} and evaluate the sensor performance by the fractional reconstruction error, Eq.~\ref{eq:error}. We average the results over 20 random training and test sets, each with 10 realizations of random noise.

\begin{figure}
\centering
\includegraphics[width=\columnwidth]{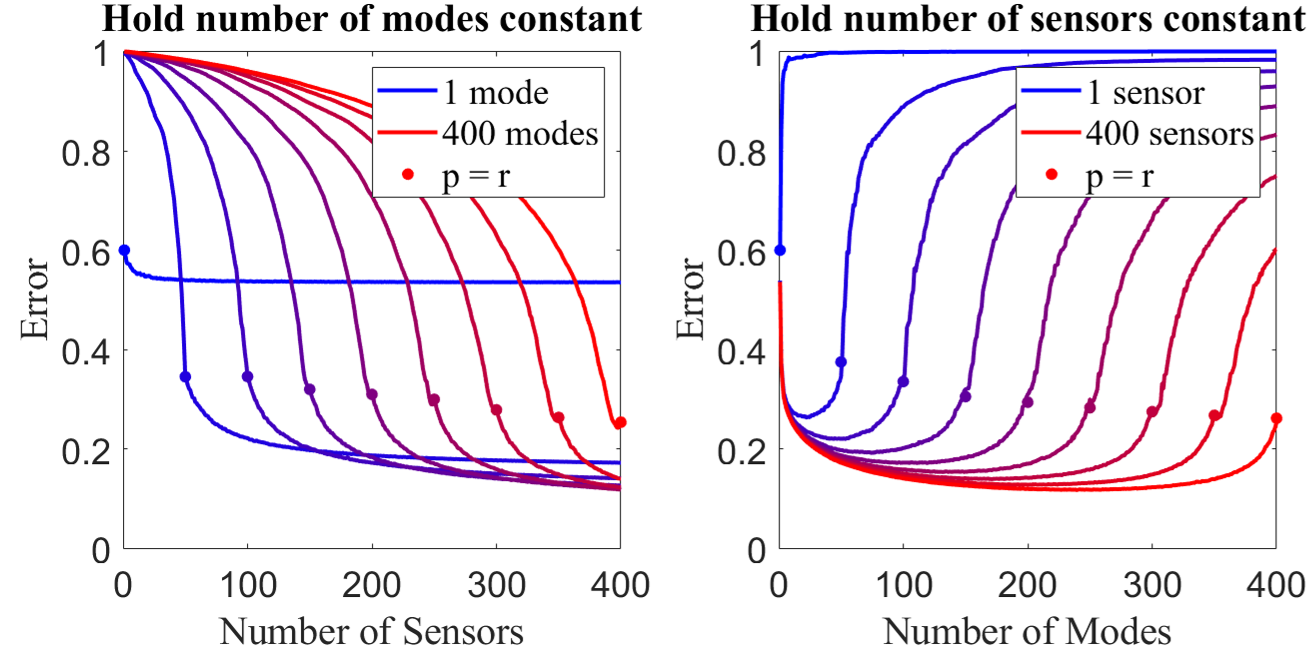}
\vspace{-.25in}
\caption{Reconstruction error for the eigenface data set, taking a basis of SVD modes. The left plot shows error versus the number of sensors as the number of modes is held constant, while the right plot gives error versus the number of modes, holding the number of sensors constant. The blue line indicates one mode (sensor), shading through to red with 400 modes (sensors) in increments of 50. The dots show the point at which the number of sensors is equal to the number of modes. Randomized oversampling is performed where $p>r$.}
\label{fig:E_FaceSVD}
\end{figure}

\begin{figure}
\centering
\includegraphics[width=\columnwidth]{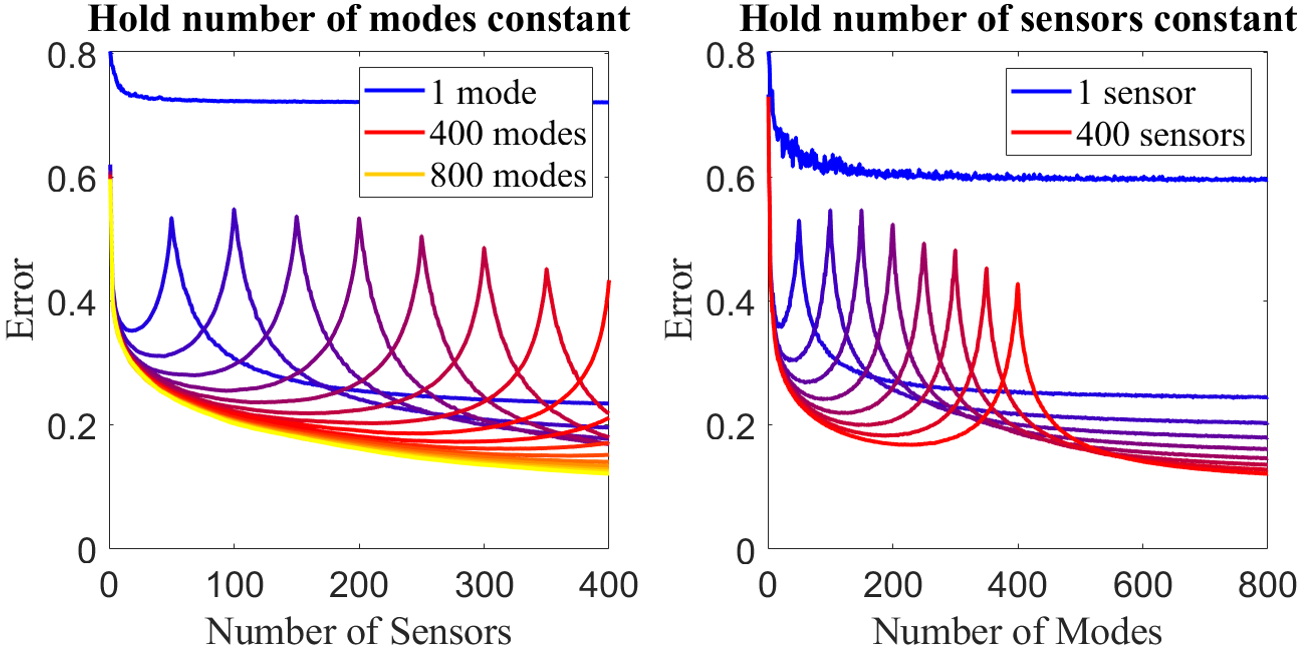}
\vspace{-.25in}
\caption{Reconstruction error for the eigenface data set, taking a basis of randomized modes, with randomized oversampling in the case of more sensors than modes. The left plot shows error versus the number of sensors as the number of modes is held constant, while the right plot gives error versus the number of modes, holding the number of sensors constant. The blue line indicates one mode (sensor), shading through to red with 400 modes (sensors), and on the left plot, continuing on to yellow with 800 modes.}
\label{fig:E_FaceRand}
\end{figure}

Fig.~\ref{fig:E_FaceSVD} and~\ref{fig:E_FaceRand} show results of varying the number of modes and sensors, with an SVD basis for the former figure and a randomized basis for the latter. Both figures have the same layout. The left plot shows reconstruction error versus the number of sensors, where each line has the same number of modes, while on the right, error is plotted as a function of the number of modes, where each line has a constant number of sensors. The blue lines have one mode (sensor), and the colors shade through to red with 400 modes (sensors). For the randomized basis, we continue through to 800 modes, shown in yellow. In the cases where $p>r$, we perform randomized oversampling, placing the first $r$ sensors with column-pivoted QR and the remaining $p-r$ sensors randomly.

Fig.~\ref{fig:E_FaceSVD} makes it clear that oversampling is indeed preferred in an SVD basis. On the left plot, the error continues to decrease as sensors are added beyond the number of modes, and equivalently on the right plot there are minima when $r<p$. Meanwhile, the lines in Fig.~\ref{fig:E_FaceRand} have pronounced peaks at $p=r$, suggesting that the randomized basis performs well with either over- or undersampling, but very poorly when the number of modes and sensors are very similar.

\begin{figure}
\vspace{-.2in}
\centering
\includegraphics[width=\columnwidth]{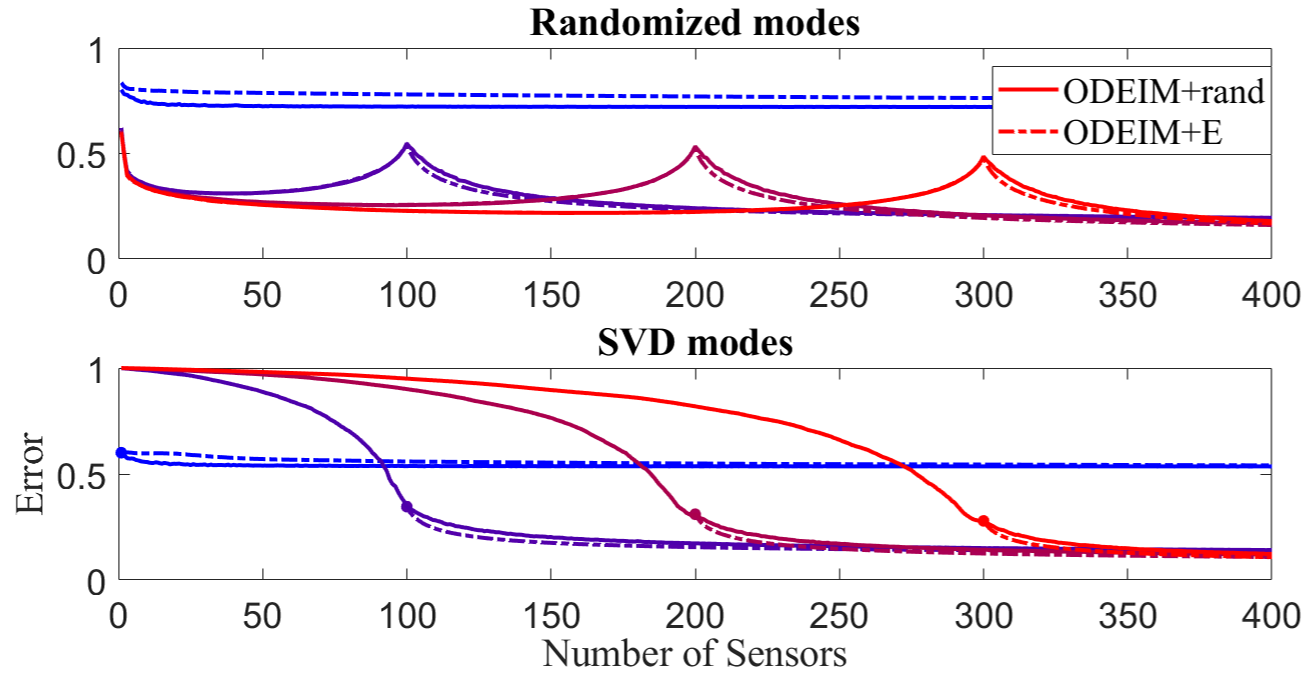}
\vspace{-.25in}
\caption{A comparison of randomized oversampling ODEIM+rand, with principled oversampling through eigenvalue tuning, ODEIM+E. In both subplots, error versus the number of sensors is plotted as the number of modes is held constant, with blue showing one mode and red showing 300 modes. Where $p>r$, randomized oversampling is shown as a solid line and principled oversampling is shown as a dashed line. The upper plot shows results with a randomized basis, and the lower plot gives results with an SVD basis.}
\label{fig:E_Face_ODEIME}
\end{figure}

\begin{figure}
\centering
\includegraphics[width=\columnwidth]{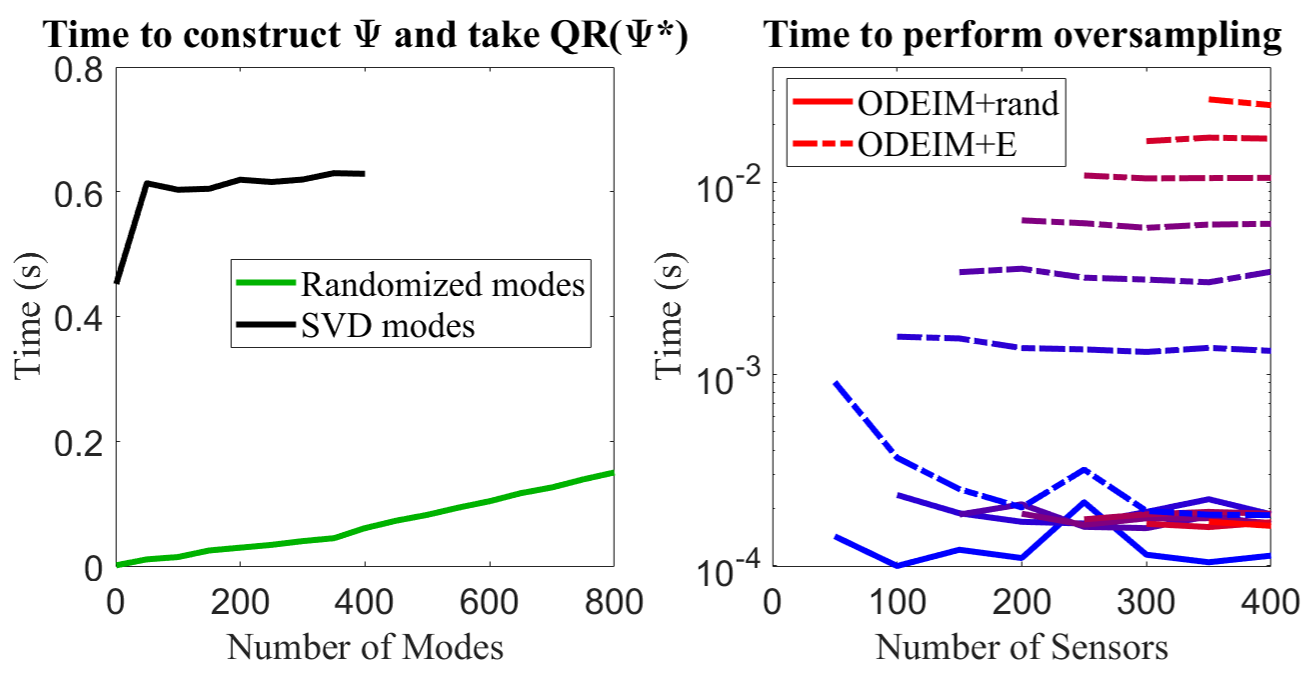}
\vspace{-.25in}
\caption{Time comparisons for basis choice (left) and oversampling method (right). For basis choice, we consider the combined time to construct the randomized and SVD bases and then to perform the column-pivoted QR decomposition. On the right, we calculate the time to perform randomized oversampling (solid lines) and principled eigenvalue-based oversampling (dashed lines), with an SVD basis. We use 1 to 300 modes and up to 400 sensors, with 1 mode plotted in blue and 300 modes in red. Both $r$ and $p$ increase in increments of 50.}
\label{fig:TimeComp}
\end{figure}

We also test principled oversampling with a method we refer to as ODEIM+E; see~\cite{peherstorfer2018stabilizing} for full details. The method places sensors so as to maximize the smallest eigenvalue of ${\bf\Theta}$, which should provide highly stable reconstructions. However, ODEIM+E is significantly slower to perform than randomized oversampling, and in this case it performs only slightly better. This can be seen in Fig.~\ref{fig:E_Face_ODEIME}, which shows reconstruction error versus $p$, with 1, 100, 200, and 300 modes. Results are shown for randomized modes in the upper plot and SVD modes in the lower, with the solid line showing randomized oversampling and the dashed line showing ODEIM+E. Principled oversampling outperforms ODEIM+rand by only about $1\%$, and given the increased runtime to perform ODEIM+E, demonstrated in Fig.~\ref{fig:TimeComp}, we determine that for the eigenface data set, it makes sense to use randomized oversampling. The right subplot of Fig.~\ref{fig:TimeComp} shows the CPU time to place the last $p-r$ sensors using both methods, plotted as a function of the number of sensors. Each line has a constant number of modes, and it is apparent that as the number of modes increases, the time to perform randomized oversampling remains constant, while the time to perform ODEIM+E increases by up to two orders of magnitude.

\begin{figure}
\vspace{-.2in}
\centering
\includegraphics[width=\columnwidth]{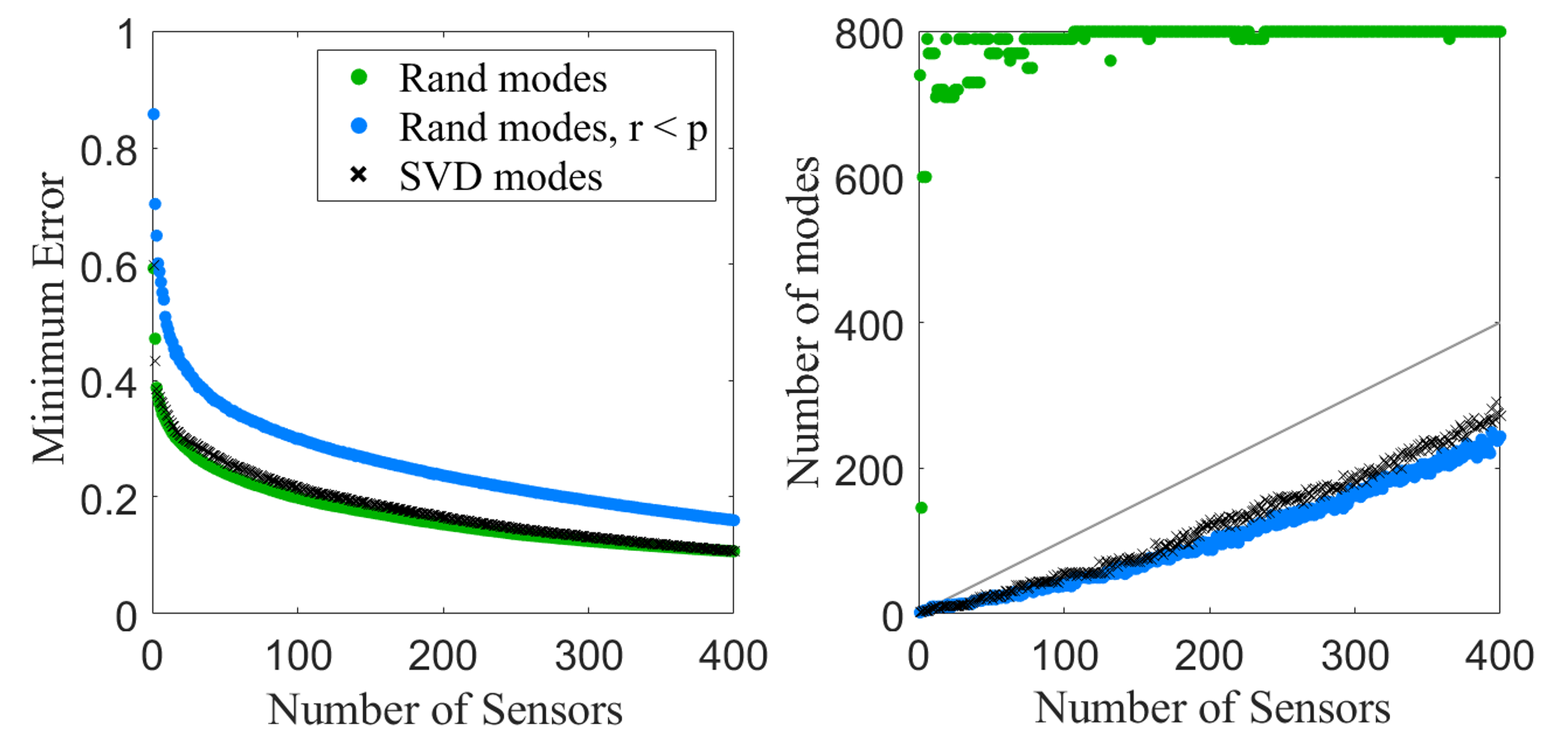}
\vspace{-.25in}
\caption{On the left is plotted the minimum possible error at a given number of sensors, for a basis of randomized modes in green, SVD modes in black, and randomized modes restricted to fewer modes than sensors in blue. The right panel shows the number of modes at which the minimum error occurs, also as a function of the number of sensors, in the same color scheme as that on the left. Also included is the $r = p$ line of number of modes and sensors being equal. No noise is incorporated here, to save on runtime.}
\label{fig:MinE_Face}
\end{figure}

Most importantly for this section, we must determine which basis provides the lowest reconstruction error. In Fig.~\ref{fig:MinE_Face}, we consider both bases and determine the lowest error at a given number of sensors, plotted on the left, and the number of modes at which this minimum error occurs, plotted on the right. We also consider the local minima present in the randomized mode plots by calculating the minimum error with $r<p$. The results indicate that the randomized and SVD bases have nearly identical minimum errors, but while the SVD basis prefers oversampling, the randomized basis requires close to the full 800 modes. When the randomized basis is restricted to oversampling, the minimum error is up to $10\%$ higher than with undersampling or taking the SVD basis. The randomized basis is faster to construct than the SVD basis, as shown on the left plot of Fig.~\ref{fig:TimeComp}, which plots the CPU time to construct ${\bf\Psi}$ and perform the column-pivoted QR decomposition as a function of the number of modes for both bases. Thus the randomized basis is appealing, but in many sensor placement examples, one of the goals is to perform rank reduction. The randomized basis only performs very well with as many modes as possible, so we decide to take an SVD basis with $p = 2r$ oversampling for the small examples used in the rest of this text. For a very large data set, the speed-up from using the randomized basis may be preferable. Note that fitting a linear regression to $p(r)$ for the SVD data yields a slope of about $p = 1.4r$. (For the randomized modes with oversampling, $p = 1.6r$.) Taking $p=2r$ ensures that the degree of oversampling is identical for any number of sensors, since sensors and modes must be integers.

\section{Multi-fidelity sensor selection}
\label{sec:Multi-fidelity}
The previous section details the underlying empirical and theoretical observations for the sensor selection problem.  In this section, we consider the central aim of the paper:  multi-fidelity sensor selection.  Specifically, cheap and expensive sensors will simply be characterized by the additive noise in a measurement.  Cheap sensors will have higher additive noise than expensive sensors.  Thus the cost of an expensive sensor improves the signal to noise ratio for the measurement.

We now allow for the presence of inhomogeneous measurement noise:
\begin{equation}
{\bf Y} = {\bf C}_{\{J\}}{\bf X}^{te} + \boldsymbol{\epsilon},
\end{equation}
where the elements of $\boldsymbol\epsilon \in \mathbb{R}^{p\times\ell}$ are given by $\epsilon_{ji} \propto \mathcal{N}(0,\sigma_j^2)$. Once the noisy measurements have been obtained, the full state reconstruction is identical to Eq.~\ref{eq:recona}. For the purposes of this paper, let the sensor noise variances take one of two values, $\sigma_j\in\{\sigma^{(exp)},\sigma^{(ch)}\}$, with $\sigma^{(exp)}$ corresponding to expensive sensors and $\sigma^{(ch)}$ to cheap sensors. This implies $\sigma^{(exp)} < \sigma^{(ch)}$.

We choose a basis of SVD modes for ${\bf\Psi}$, which provides nearly identical reconstruction errors to randomized rank reduction but with far fewer modes. We use $r = p/2$ modes and randomized oversampling, which is orders of magnitude faster than principled oversampling and performs only slightly less well. Therefore, to place a total of $p$ sensors, we perform the column-pivoted QR decomposition on the first $p/2$ modes and use the first $p/2$ pivots, then randomly place the remaining $p/2$ sensors. Now we must further consider two options for placing $p^{(exp)}$ expensive sensors and $p^{(ch)}$ cheap sensors (where $p^{(exp)} + p^{(ch)} = p$). The expensive sensors could be placed at the first $p^{(exp)}$ locations:
\begin{equation}
\sigma_{1:p^{(exp)}} = \sigma^{(exp)}, \hspace{12pt} \sigma_{p^{(exp)}+1:p} = \sigma^{(ch)},
\label{eq:ExpFirst}
\end{equation}
or the expensive sensors could be placed on the last set of selected locations:
\begin{equation}
\sigma_{1:p^{(ch)}} = \sigma^{(ch)}, \hspace{12pt} \sigma_{p^{(ch)}+1:p} = \sigma^{(exp)}.
\label{eq:ExpLast}
\end{equation}
There is logic behind either choice. Since the QR algorithm selects pivots in order of approximate importance, measuring the first set of selected locations with high accuracy should lead to better overall reconstructions. However, locations are chosen to iteratively maximize the variance of the basis modes, so that the final set of locations, although randomly chosen, almost certainly has a smaller variance than the first set. Accurately measuring the last set of locations means better resolving fine details on the reconstruction, while there is a smaller signal-to-noise ratio on the high-variance first set of sensors, meaning that even cheap noisy sensors should be able to capture these modes relatively accurately.

The choice of where to place the expensive sensors depends on the parameters of the particular problem, but we find that generally the performance is better when the expensive sensors are placed on the first set of locations, following Eq.~\ref{eq:ExpFirst}. In the asymptotic cases we will consider below, it will be better to use just one type of sensor, so the choice is less important. Therefore, for clarity, we only show results from placing expensive sensors on the first set of sensor locations.

The parameter space for the multi-fidelity sensor selection problem also includes the cheap and expensive sensor noise levels $\sigma^{(ch)}$ and $\sigma^{(exp)}$, and the costs of both types of sensor, $c^{(ch)}$ and $c^{(exp)}$. We assume there is a set budget $B$, such that
\begin{equation}
c^{(ch)} p^{(ch)} + c^{(exp)} p^{(exp)} \le B.
\label{eq:Budget}
\end{equation}
We find that depending on all of these factors, it may be best to use all cheap sensors, all expensive sensors, or a mixture of both. We will explore different parameter regimes and their effect on what type of sensor to use and where to place them in the next section.

\section{Multi-fidelity sensor results}
\label{sec:Results}
Because the parameter space is combinatorially large, we consider only a few asymptotic regimes and determine in these cases whether all cheap or all expensive sensors perform better. The cases we explore fall into one of nine categories comprising combinations of very low and very high $\sigma^{(exp)}$, $\sigma^{(ch)}$, $p^{(exp)}$, and $p^{(ch)}$.

\begin{figure}
\centering
\includegraphics[width = \columnwidth]{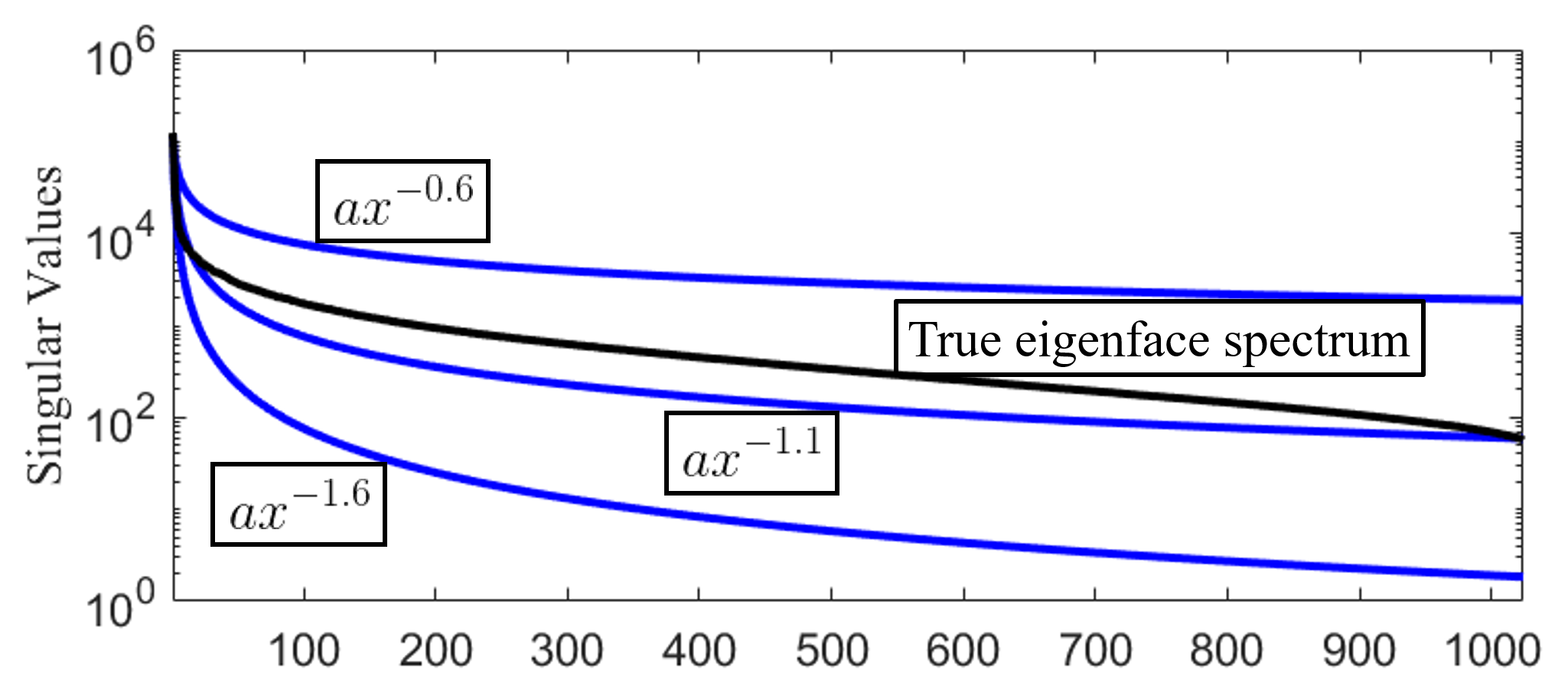}
\vspace{-.25in}
\caption{The singular values of the eigenface data set, plotted in black, and the artificial singular value spectra we use in this section, shown in blue.}
\label{fig:SingValues}
\end{figure}

We also wish to account for different kinds of data, so we construct three data sets with slow, medium, and fast singular value decays. We do this by using the left and right singular vectors from the Yale B Face data set combined with artificial singular values. We fit a power law function, $y = ax^b$, to the true singular values of the eigenface data set and find that they have the approximate form
\begin{equation}
y = (1.21\times10^5)x^{-1.14}.
\end{equation}
We assume that the exponent $b = -1.14$ corresponds to medium singular value decay, and construct the artificial data sets with $b = -0.6,-1.1,$ and $-1.6$ (we retain $a = 1.21\times10^5$). The singular values are plotted in Fig.~\ref{fig:SingValues}. The data set with $b = -1.6$ has low rank, with $90\%$ of the system's energy being captured by just 23 modes. The data set with $b=-1.1$ needs 355 modes, while the data set with $b=-1.6$ is high rank, requiring 798 modes to capture $90\%$ of its energy.

\begin{figure}
\centering
\includegraphics[width=\columnwidth]{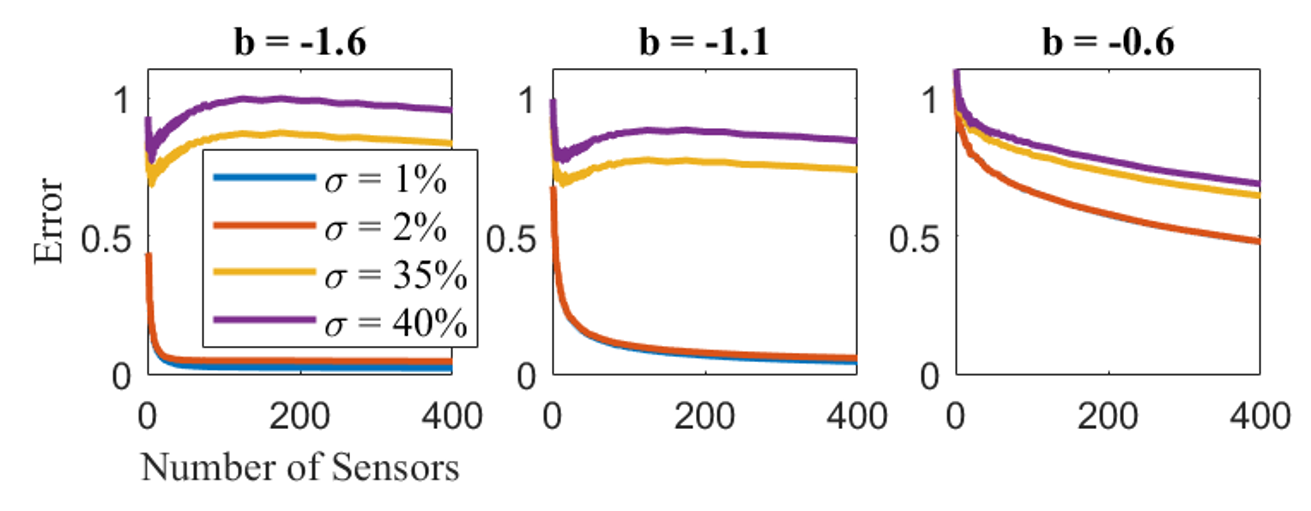}
\vspace{-.25in}
\caption{Reconstruction error versus number of sensors for three data sets with three different singular value exponents $b$. Results are given with just one type of sensor at a time, with four different noise levels, two very low and two very high.}
\label{fig:AllOneType}
\end{figure}

For reference, reconstruction results with one type of sensor are shown for the three artificial data sets in Fig.~\ref{fig:AllOneType}. Each subplot shows reconstruction error plotted against the number of sensors for one of the three data sets. We consider two high and two low noise levels. When $\sigma = 1\%$ and $2\%$, the data sets with fast and medium decaying singular values have very low reconstruction errors. For both high noise level cases, these data sets have very high errors and exhibit noise amplification, where the error increases with the number of sensors, despite performing oversampling. The $b=-0.6$ data set always has high reconstruction errors, but the error always decreases as more sensors are added.

\begin{figure*}
\vspace{-.15in}
\hspace{-24pt}\includegraphics[width = 1.1\textwidth]{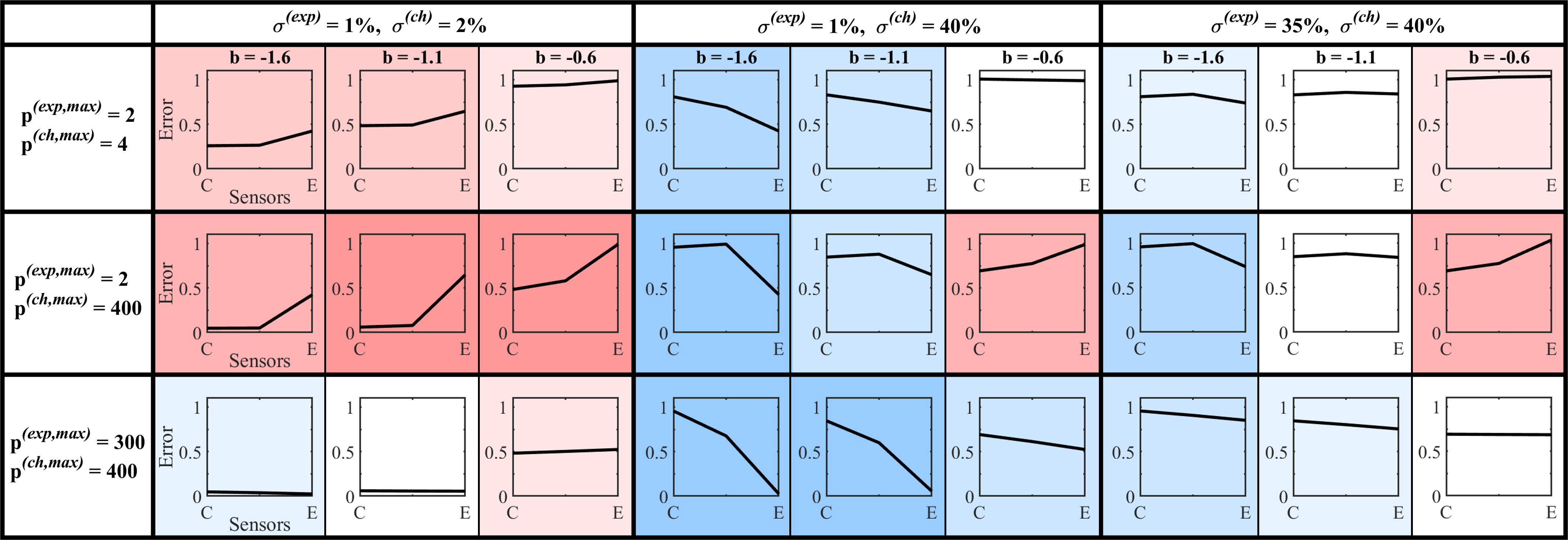}
\vspace{-.25in}
\caption{Results for placing two different types of sensors in the nine asymptotic regimes of cheap and expensive sensor number and noise level. The plots show reconstruction error versus the number of cheap and expensive sensors, with the first data points (marked as C on the $x$-axis) using all cheap sensors and the last (E on the $x$-axis) using all expensive sensors. The plots are color coded based on whether error is lower using all cheap sensors (red), all expensive sensors (blue), or if the results are inconclusive (white, errors are within 2\% of each other), as well as shaded from light to dark based on the magnitude of the difference.}
\label{fig:ResultsChart}
\end{figure*}

The main results of this paper are shown in Fig.~\ref{fig:ResultsChart}. The figure is split into nine sections, where the left column has low noise ($\sigma^{(exp)} = 1\%$, $\sigma^{(ch)} = 2\%$), the middle column has low $\sigma^{(exp)} = 1\%$ and high $\sigma^{(ch)} = 40\%$, and the right column has high-noise sensors, with $\sigma^{(exp)} = 35\%$ and $\sigma^{(ch)} = 40\%$. The top row has a small number of maximum allowed sensors, 2 and 4 for expensive and cheap, respectively. The middle row uses a maximum of 2 expensive and 400 cheap sensors, and the bottom row has a maximum of 300 expensive and 400 cheap sensors. In each section, results from the three data sets are plotted separately. From left to right they have $b = -1.6$, $-1.1$, and $-0.6$. Each of the panels shows reconstruction error as the proportion of cheap and expensive sensors is varied. Along the $x$-axis, the number of expensive sensors increases from zero to $p^{(exp,max)}$ (marked ``E"), while the number of cheap sensors decreases from $p^{(ch,max)}$ (marked ``C") to zero. At the midpoint, a mix of cheap and expensive sensors is used. All combinations satisfy Eq.~\ref{eq:Budget}, where the budget and sensor costs are adjusted to satisfy both set values of $p^{(exp,max)}$ and $p^{(ch,max)}$ for every row. Results are averaged over 20 randomized training and test sets, each with 20 cross validations over sensor location and 10 noise realizations.

If the error is lower at point ``C", then it is better to use all cheap sensors and the plot is colored red. Blue plots have lower reconstruction errors at ``E", meaning that it is preferable to use all expensive sensors. The plots are shaded from dark to light, with dark indicating that there is a large difference in reconstruction error between the two types of sensors, and light meaning that the reconstruction errors are similar. Finally, plots are colored white when there is less than a $2\%$ difference in using all cheap and all expensive sensors.

Although we have not yet found a rule for when to use all cheap versus all expensive sensors, the figure does reveal some trends. When both noise levels are very low, it is almost always better to use a large number of cheap sensors. When $\sigma^{(exp)}$ is low and $\sigma^{(ch)}$ is high, or when both sensor types have high noise levels, it is more often better to choose a small number of expensive sensors.

There are exceptions to the general trends, and it is clear that the rank of the data set plays an important role in the results. The low-rank system with $b=-1.6$ is more likely to perform better with a small number of expensive sensors, while the high-rank data with $b=-0.6$ usually does better with a larger number of cheap sensors, even when those cheap sensors have a high noise level. The system with medium singular value decay is the most likely to have comparable performances with all cheap and all expensive sensors.



In these asymptotic regimes, it is never preferable to have a mix of cheap and expensive sensors, and there are a few cases where a mixture has the worst performance, such as the $b=-1.6$ panel of the middle row and column. Apparently the balance between improving results by adding sensors and improving results by reducing noise is suboptimal in this case. 

We note that many of these reconstructions are very poor no matter which sensors are used. An eigenface reconstruction with error greater than around $20\%$ begins to look unrecognizable, so the high-noise or low-sensor cases where the reconstruction errors approach or exceed $100\%$ are entirely dominated by noise. In a real-world case like this, the engineer or scientist would need to either purchase a larger number of expensive sensors or design sensors with less noise.

\begin{figure*}
\centering
\includegraphics[width = 0.8\textwidth]{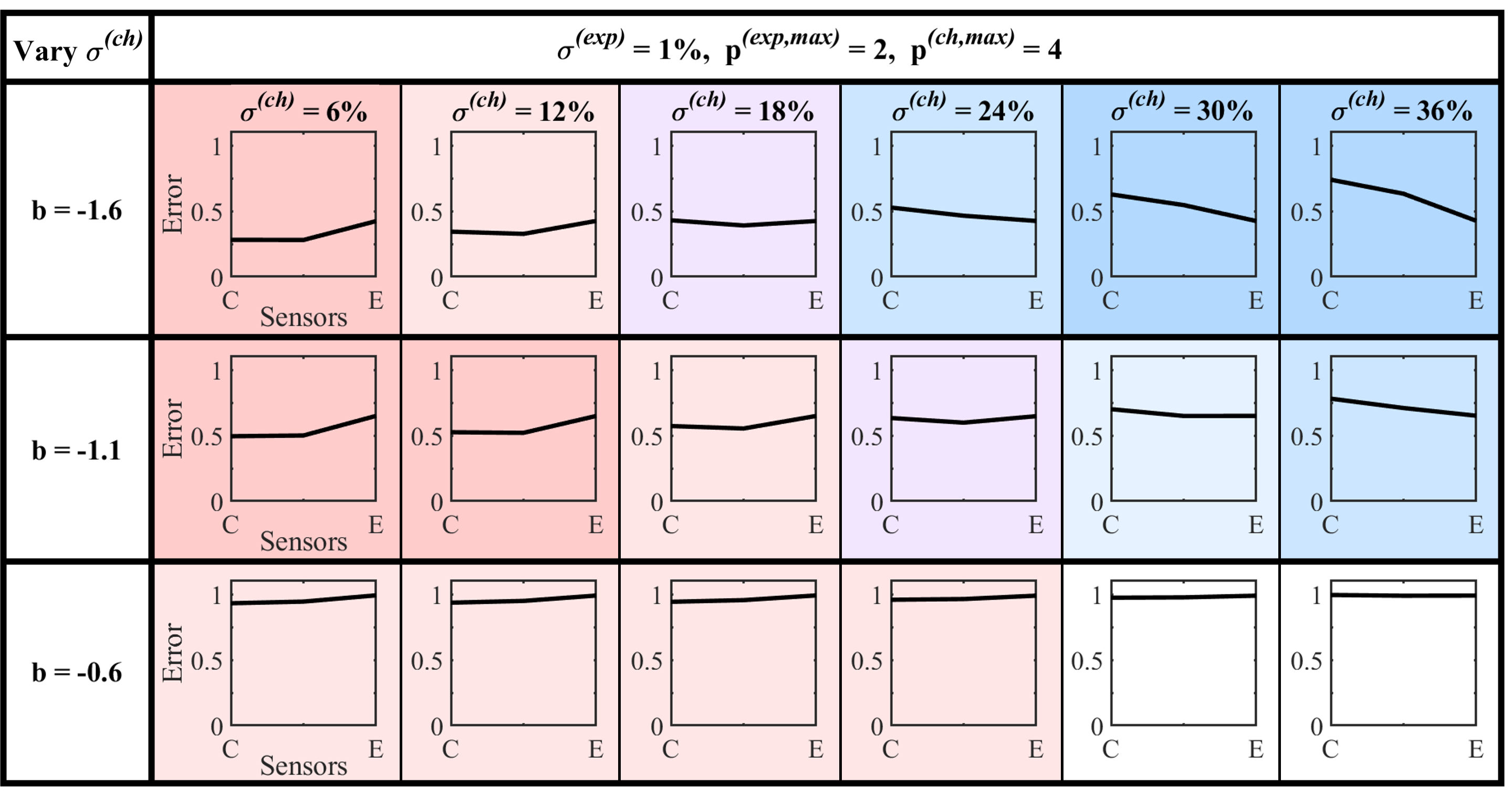}
\vspace{-.1in}
\caption{Slices across cheap sensor noise level. Each row corresponds to one singular value exponent and each column has a different value of $\sigma^{(ch)}$, beginning at $6\%$ and increasing to $36\%$. All panels are in the regime of low $\sigma^{(exp)}$, $p^{(exp)}$, and $p^{(ch)}$. As the results transition from favoring all cheap sensors to favoring all expensive sensors, the data sets with $b = -1.6$ and $-1.1$ pass through a point where it is optimal to have a mix of one expensive sensor and two cheap sensors, shown in purple. The rest of the color scheme is as in Fig.~\ref{fig:ResultsChart}.}
\label{fig:ResultsSliceSigc}
\end{figure*}

\begin{figure*}
\vspace{-.1in}
\centering
\includegraphics[width = 0.85\textwidth]{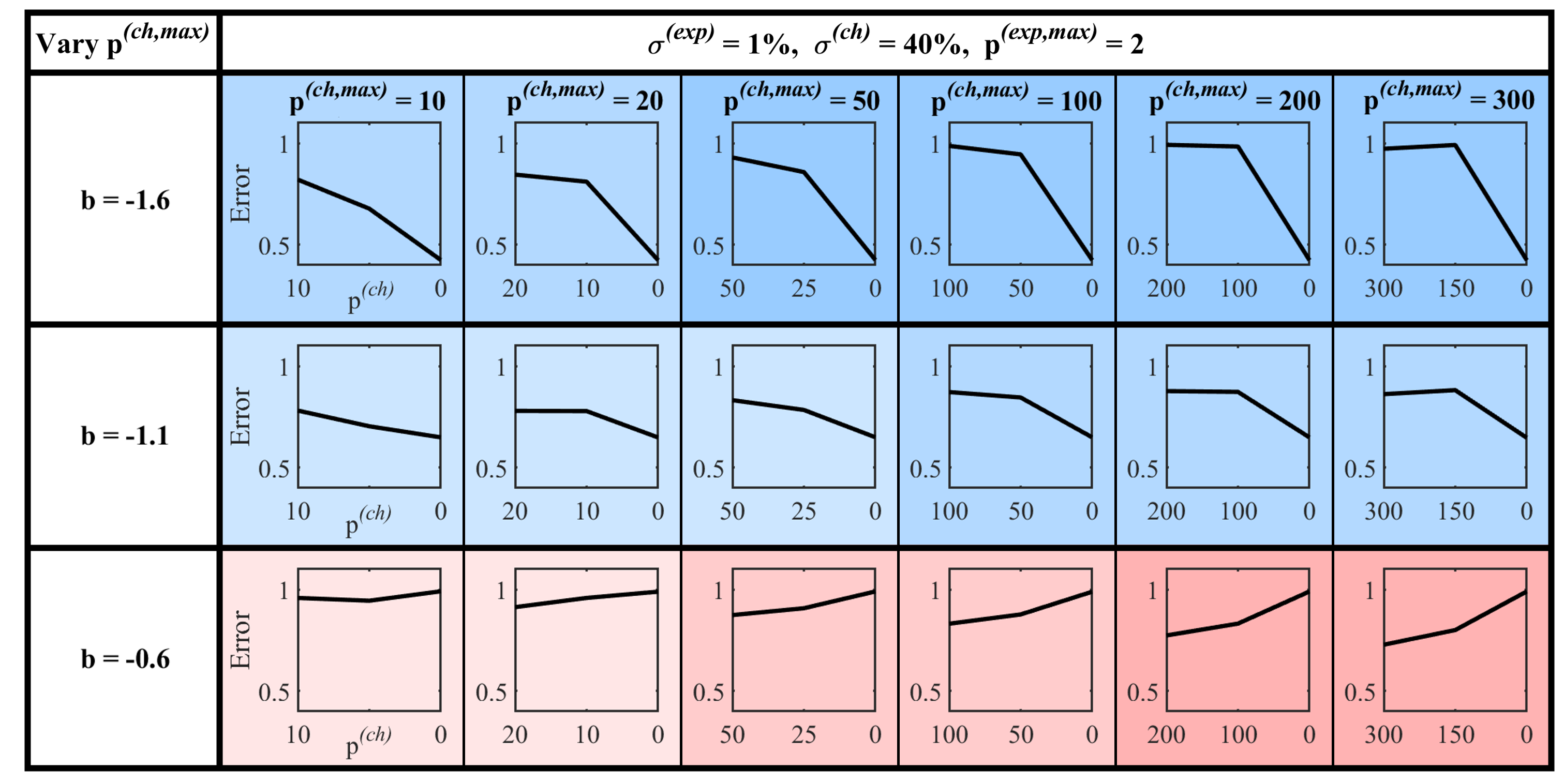}
\vspace{-.1in}
\caption{Slices across maximum number of cheap sensors. As in Fig.~\ref{fig:ResultsSliceSigc}, each row corresponds to a value of $b$, but here each column has a different maximum number of cheap sensors, increasing from left to right, equivalent to decreasing $c^{(ch)}$. In this case, $p^{(ch)}$ is labeled on the $x$-axis, but with $p^{(exp,max)}=2,$ $p^{(exp)}$ implicitly increases along the $x$-axis from zero to two. All panels have $\sigma^{(exp)} = 1\%$ and $\sigma^{(ch)} = 40\%$.}
\label{fig:ResultsSlicePc}
\end{figure*}

\begin{figure*}
\centering
\includegraphics[width = 0.8\textwidth]{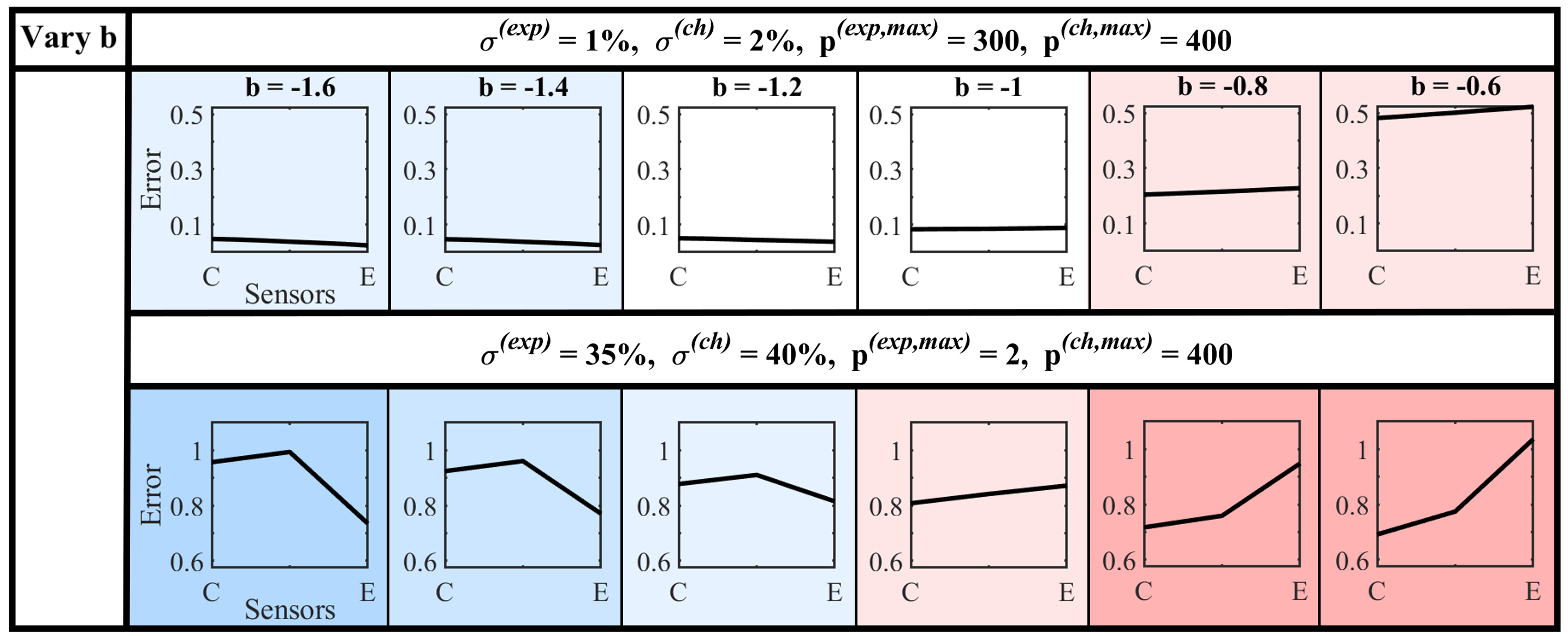}
\vspace{-.1in}
\caption{Slices across singular value exponent. Each column has a different value of $b$, increasing from $-1.6$ to $-0.6$. The top row is in the regime of low $\sigma^{(ch)}$ and $\sigma^{(exp)}$ and high $p^{(ch)}$ and $p^{(exp)}$. The lower row has high noise values, with low $p^{(exp)}$ and high $p^{(ch)}$.}
\label{fig:ResultsSliceExp}
\end{figure*}

\begin{figure}
\vspace{-.1in}
\centering
\includegraphics[width = 0.9\columnwidth]{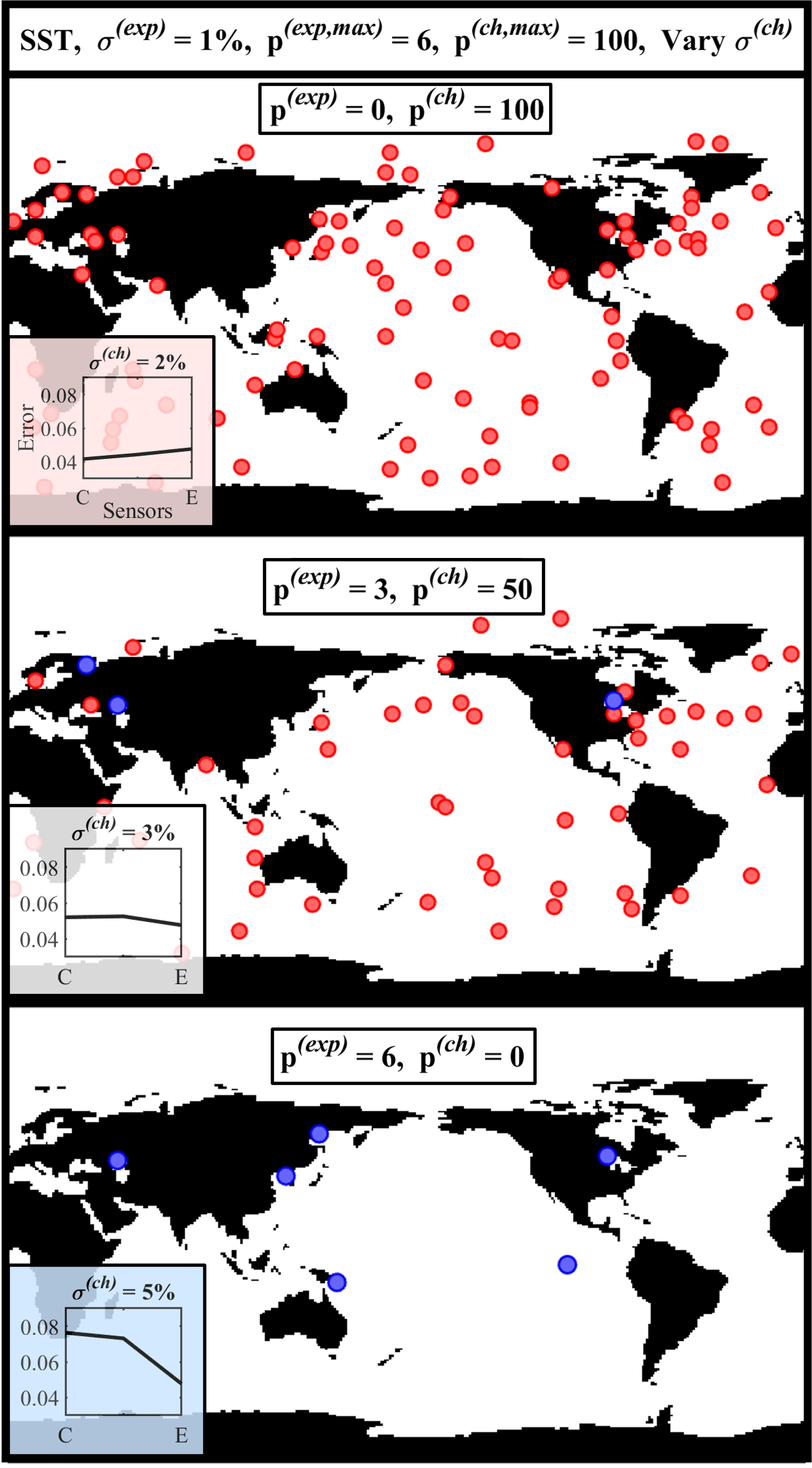}
\vspace{-.1in}
\caption{Sensor location results for the sea surface temperature data set, selecting up to six expensive sensors with 1\% noise level and up to 100 cheap sensors with varying noise levels. The cheap sensors are shown in red and the expensive sensors in blue. The insets give the error versus the number of cheap and expensive sensors, as in previous figures.  The trade-off between cheap and expensive sensors, and their combined use, is highly informative for the global monitoring of the sea surface temperature.}  
\label{fig:ResultsSST}
\end{figure}


As previously discussed, the parameter space for the multi-fidelity sensor problem is very large, and exploring all cases between the asymptotic regimes shown in Fig.~\ref{fig:ResultsChart} is prohibitive. However, in Fig.~\ref{fig:ResultsSliceSigc},~\ref{fig:ResultsSlicePc}, and~\ref{fig:ResultsSliceExp}, we show a few intermediary results, made by varying one parameter at a time.

Fig.~\ref{fig:ResultsSliceSigc} shows results for all three singular value exponents, holding $\sigma^{(exp)}$, $p^{(exp,max)}$, and $p^{(ch,max)}$ constant at low values, and increasing $\sigma^{(ch)}$ from a low to a high value. The data sets with fast and medium singular value decays both transition from favoring all cheap sensors to all expensive sensors, passing through a small window of $\sigma^{(ch)}$ values where a mix of sensor types produces the best results (shown in purple). A mix does not outperform all cheap or all expensive sensors by a large amount, only by about 3 or 4\%. The data set with slow singular value decay slightly favors the use of all cheap sensors at low values of $\sigma^{(ch)}$ and transitions to seeing similar results with either type of sensor. There is no point where it is significantly better to use a mix, and since the system has a very high rank, all of the reconstructions are very poor.

We vary the maximum number of cheap sensors, with $p^{(exp,max)} = 2$, $\sigma^{(exp)} = 1\%$, and $\sigma^{(ch)} = 40\%$, with results shown for all three values of $b$ in Fig.~\ref{fig:ResultsSlicePc}. In this case, increasing the maximum number of cheap, noisy sensors intensifies the results seen in this regime in Fig.~\ref{fig:ResultsChart}. For the $b=-1.6$ and $-1.1$ systems, adding additional, highly noisy sensors decreases the reconstruction quality, and for $p^{(ch,max)} = 300$, a mix of sensor types yields the worst performance. Meanwhile, for the high-rank system, adding sensors improves the reconstruction, even though the cheap sensors have extremely high noise levels. 
Based on the parameters chosen here, none of the reconstructions are very good (notice the range of the $y$-axes) for any of the data sets.

Finally, we vary the singular value exponent $b$ for two different sets of the remaining parameters, as shown in Fig.~\ref{fig:ResultsSliceExp}. In both cases, the systems transition from preferring a small number of expensive sensors to a large number of cheap sensors as $b$ is increased from $-1.6$ to $-0.6$. This makes sense, as it only requires a small number of sensors to produce a good reconstruction of a low-rank system, while adding sensors to a high-rank system almost always improves the reconstruction, even when the sensors have high noise levels. As no point is it better to have a mix of sensor types, and when both sensor types have high noise levels and the system is low-rank, a mix gives the poorest results.

To demonstrate that the trends above also apply to real-world data, we consider the NOAA weekly sea surface temperature (SST) data~\cite{noaa2018, banzon2016, reynolds2007}. This data set consists of weekly sea surface temperature snapshots from 1990 to 2016, on a $360\times180$ grid. We select sensors as above, using an SVD basis with $p=2r$ oversampling in the case of $p>10$, where noise amplification begins to be significant (when $p<10$, we set $r=p$). Expensive sensors have $1\%$ noise level and $p^{(exp,max)} = 6$, and we allow up to 100 cheap sensors with noise levels of $2$, $3$, and $5\%$. Results are given in Fig.~\ref{fig:ResultsSST}, where the panels show example sensor placements corresponding to optimum performance given the conditions: 100 cheap sensors when $\sigma^{(ch)}$ is low, and 6 expensive sensors in the case of a relatively high $\sigma^{(ch)}$. When $\sigma^{(ch)} = 3\%$, the performances are comparable for each distribution of sensors, so for interest we plot the locations of a mix of 3 expensive and 50 cheap sensors. The insets show error versus the number of cheap and expensive sensors, as in previous figures. We find that the expensive sensors are more likely to be placed in landlocked seas or lakes, as these measurements are fairly independent from the oceans. Finally, recall that with a mix of sensors, the expensive sensors were placed at the first QR pivots. However, we find that placing the first $p/2 = 27$ cheap sensors at the principled QR pivots and the remaining 26 cheap and expensive sensors randomly gave similar results on average.

All of these figures cover only a small section of the total parameter space for the multi-fidelity sensor selection problem. We have discovered trends in the results, but no analytic rule for when to choose which type of sensor. Future work must continue to investigate the influence of sensor noise levels, costs, budget, and system rank on reconstruction quality and the choice of sensor type.

\section{Discussion}
\label{sec:Conclusion}
We consider sparse, multi-fidelity sensor selection for full-state reconstruction in the case of two types of available sensors:  (i) low noise with high cost (large signal-to-noise) and (ii) high noise with low cost (low signal-to-noise). The problem and results are complex and nuanced, with a large, non-convex parameter space.  Regardless, we provide an initial exploration of a few asymptotic cases of low and high noise levels, and low and high numbers of sensors. Because results are dependent on the rank of the measured data, we construct three artificial data sets with slow, medium, and fast singular value decay. We employ the column-pivoted QR decomposition for sensor placement and find a few general trends for the multi-fidelity sensor selection. Under our chosen asymptotic conditions, it is never better to use a mix of both types of sensors. If both sensors have low noise, it is usually better to use a large number of cheap sensors than a small number of expensive sensors. If the cheap sensors have much higher noise levels than the expensive sensors, then a small number of expensive sensors usually performs better. And if both types of sensor have very high noise levels, reconstructions are generally very poor and which type performs better is rank-dependent: low-rank systems slightly favor a small number of expensive sensors, while high-rank systems do slightly better with a large number of cheap sensors.

We also test the effect of over- and under-sampling with one type of sensor given one of two common basis choices. With an SVD basis, oversampling stabilizes the reconstruction, while with a randomized reduced basis, taking more modes than sensors captures more of the system's energy and leads to improved results. We find that for the Yale B Face data set, the SVD with oversampling yields nearly identical reconstructions to randomized modes with under-sampling, but while the SVD performs best with only moderate oversampling, the randomized basis results improve continually as more modes are added. In most applications, sparse sensor placement is performed alongside rank reduction, so the SVD will probably be the preferred choice. However, randomized rank reduction is faster to perform, so it may be preferable for a very large system.

This has not been a complete treatment of the multi-fidelity sensor placement problem. The ultimate objective is to discover a set of principles for how many of each sensor type to use and where to place them, given the rank of the data, the sensor costs and noise levels, and a set budget.  Such principles are difficult to discover except in various asymptotic regimes.  Regardless, the work suggests how further studies can be used to reveal more trends, even in non-asymptotic cases. Further extensions could include a weighted or statistical reconstruction method, which could help account for sensor noise, improving reconstructions and perhaps making a mix of sensor types more viable. It would also be interesting to consider more than two types of sensors, or sensors that are multi-fidelity in some other way, like modality, bandwidth, or time resolution. We could also improve cost effectiveness by incorporating a cost function into the column-pivoted QR algorithm, as in~\cite{clark2018greedy}. Sensor placement is often treated simplistically mathematically, but all the practicalities of physical sensors, like cost and noise level, make it a highly complex problem. This text treats just one facet of the practical problem, but even so, it should lead to further developments, better reconstructions, and cost savings for practitioners looking to deploy networks of sensors.

\section*{Acknowledgment}
E. Clark and S. L. Brunton acknowledge support from the Boeing Company.
S. L. Brunton acknowledges support from the Air Force Office of Scientific Research (FA9550-18-1-0200).
J. N. Kutz acknowledges support from the Air Force Office of Scientific Research (FA9550-19-1-0011).

\ifCLASSOPTIONcaptionsoff
  \newpage
\fi








%



\begin{IEEEbiography}[{\includegraphics[width=1in]{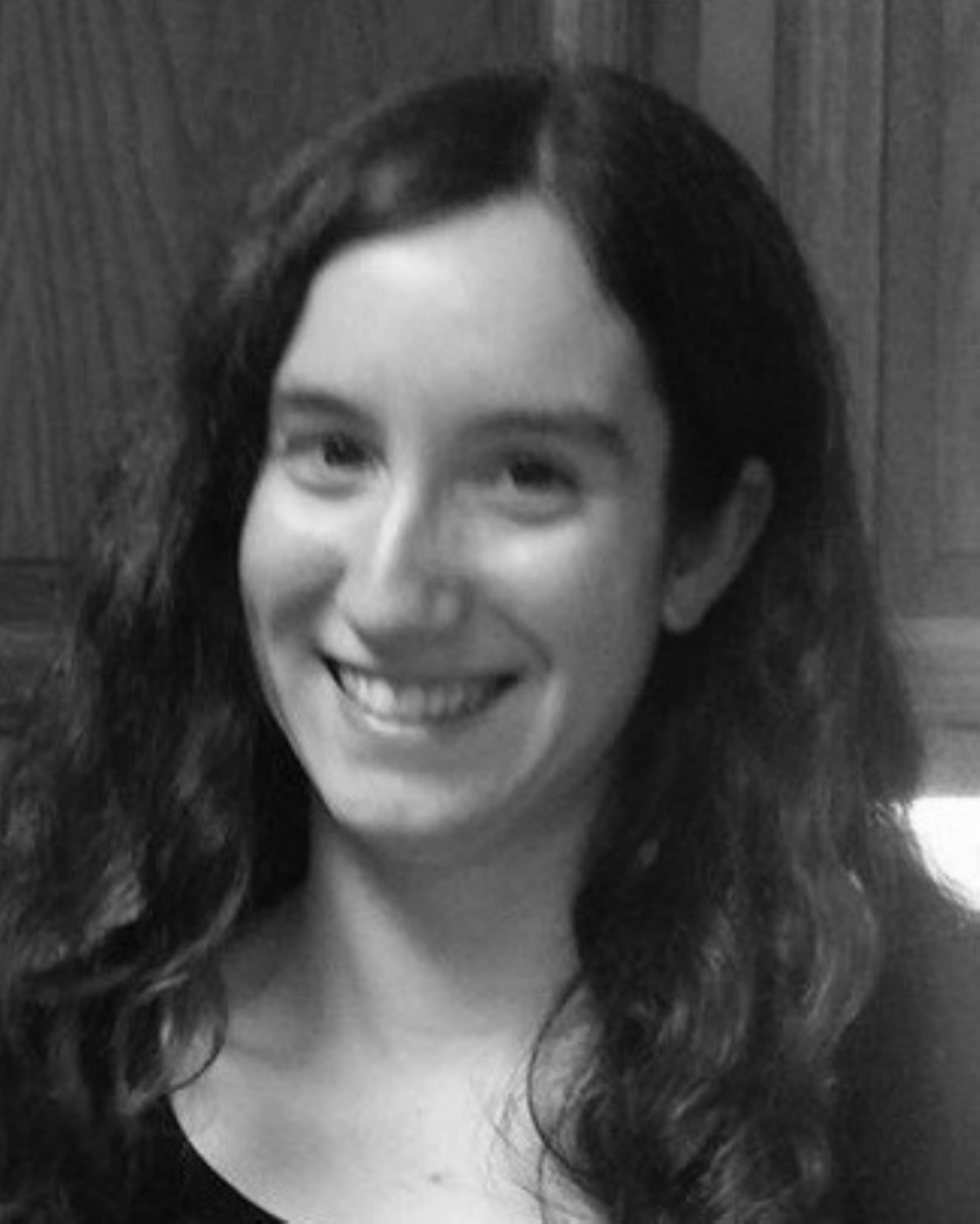}}]{Emily Clark}
received the B.S. degree in physics from Bates College, Lewiston, ME, in 2015. She is a Ph.D. candidate of physics at the University of Washington.
\end{IEEEbiography}

\begin{IEEEbiography}[{\includegraphics[width=1in]{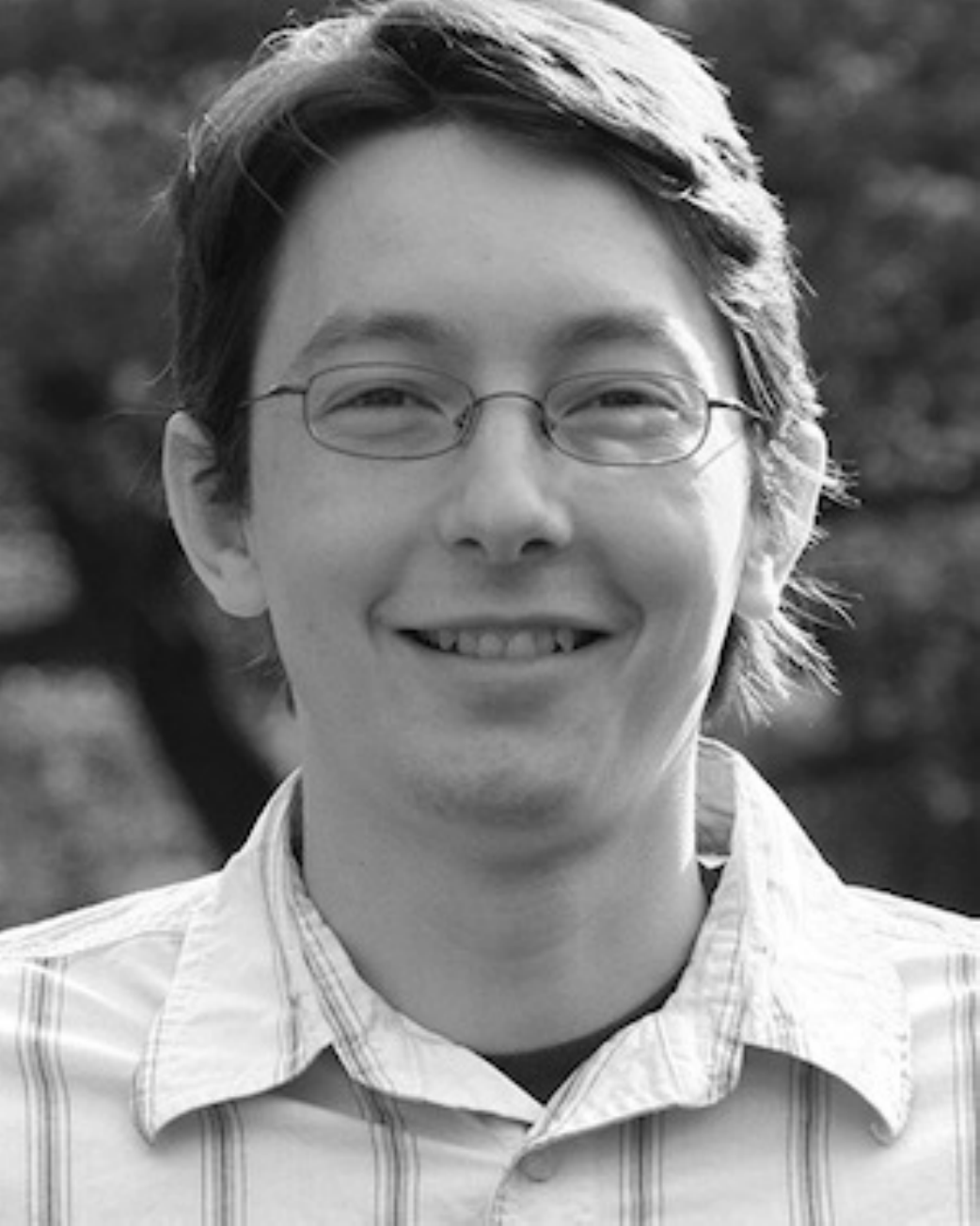}}]{Steven~L.~Brunton}
(Senior Member, IEEE) received the B.S. degree in mathematics with a minor 
in control and dynamical systems from the California 
Institute of Technology, Pasadena, CA, in 2006, and 
the Ph.D. degree in mechanical and aerospace engineering 
from Princeton University, Princeton, NJ, in 2012.  He 
is an Associate Professor of mechanical engineering and 
a data science fellow with the eScience institute at the 
University of Washington.
\end{IEEEbiography}

\begin{IEEEbiography}[{\includegraphics[width=1in]{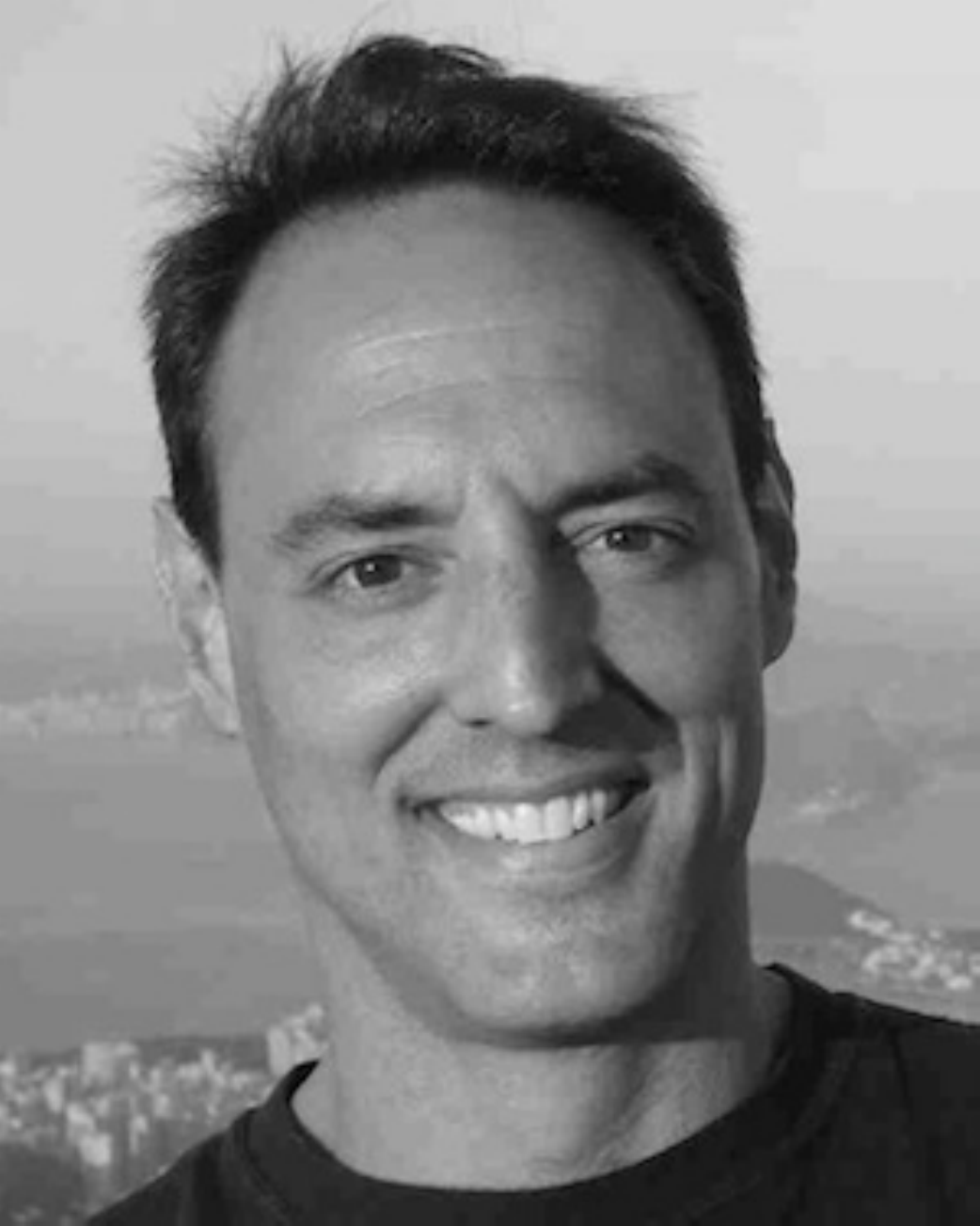}}]{J.~Nathan Kutz}
(Member, IEEE) received the B.S. degrees in physics and mathematics 
from the University of Washington, Seattle, WA, in 1990, 
and the Ph.D. degree in applied mathematics from 
Northwestern University, Evanston, IL, in 1994.  He is 
currently a Professor of applied mathematics, adjunct 
professor of physics, mechanical engineering, and electrical engineering, and a 
senior data science fellow with the eScience institute 
at the University of Washington.
\end{IEEEbiography}

%







\end{document}